\documentclass[fleqn,usenatbib]{mnras}

\usepackage{newtxtext,newtxmath}

\usepackage[T1]{fontenc}

\DeclareRobustCommand{\VAN}[3]{#2}
\let\VANthebibliography\thebibliography
\def\thebibliography{\DeclareRobustCommand{\VAN}[3]{##3}\VANthebibliography}

\usepackage{graphicx}
\usepackage[utf8]{inputenc}
\usepackage{hyperref}
\hypersetup{
    colorlinks=true,
    linkcolor=blue,
    citecolor=blue,
    filecolor=magenta,      
    urlcolor=cyan,
    pdfpagemode=FullScreen,
    }
    
\usepackage{dcolumn}
\usepackage{bm}
\usepackage{siunitx}
\usepackage{booktabs}
\usepackage{float}
\usepackage{amsmath}
\usepackage{subfig}
\usepackage{caption}
\usepackage{threeparttable}
\usepackage{longtable}
\usepackage{longtable}
\usepackage{booktabs}
\usepackage{pdflscape}
\usepackage{tabularx}
\usepackage{multirow}
\usepackage{geometry}
\usepackage{appendix}
\usepackage{pdflscape}

\newcommand{\ciii}{C\,\textsc{iii}]}
\newcommand{\lya}{Ly$\alpha$}
\newcommand{\heii}{He\,\textsc{ii}}
\newcommand{\kms}{km\,s$^{-1}$}

 \title[\ciii\ as a tracer of \lya\ at $z>6$]{C\,\textsc{iii}]\,$\lambda1909$ emission as an alternative to Ly$\alpha$ in the reionization era: the dependence of C\,\textsc{iii}] and Ly$\alpha$ at $3<z<4$ from the VANDELS survey}

\author[M. H. Cunningham et al.]{M. H. Cunningham,$^{1}$\thanks{mark.cunningham.20@ucl.ac.uk} A. Saxena,$^{2,1}$ R. S. Ellis$^{1}$ and L. Pentericci$^{3}$
\\
$^{1}$ Department of Physics and Astronomy, University College London, London, WC1E 6BT, UK\\
$^{2}$ Department of Physics, University of Oxford, Denys Wilkinson Building, Keble Road, Oxford OX1 3RH, UK \\
$^{3}$ INAF - Osservatorio Astronomico di Roma, Via di Frascati 33, 00078, Monte Porzio Catone, Italy
}

\date{Accepted XXX. Received YYY; in original form ZZZ}

\pubyear{2023}

\begin{document}
\label{firstpage}
\pagerange{\pageref{firstpage}--\pageref{lastpage}}
\maketitle

\begin{abstract}
The velocity offset of Ly$\alpha$ emission from a galaxy's systemic redshift is an excellent tracer of conditions that enable the escape of Ly$\alpha$ photons from the galaxy, and potentially the all-important hydrogen ionizing Lyman continuum photons. However at $z\geq6$, Ly$\alpha$ is often heavily attenuated by the neutral intergalactic medium. Here we investigate the utility of C\textsc{iii}]\,$\lambda\lambda1907,1909$ emission, usually the brightest UV line after Ly$\alpha$, as a proxy estimating the Ly$\alpha$ velocity offset ($\Delta v_{\rm{Ly}\alpha}$). To do so, we use analogues of reionization era galaxies based upon 52 star-forming galaxies with robust C\,\textsc{iii}] detections drawn from the VANDELS survey. Our sample spans a broad UV magnitude range of $-18.5 < M_{\rm{UV}} < -22.0$, with a sample average value of EW(C\,\textsc{iii}]) $=5.3$\,\AA. We find a slight increase of EW(C\,\textsc{iii}]) with increasing EW(Ly$\alpha$), but find a large range of EW(C\,\textsc{iii}]) $\sim$1-13\,\AA\ particularly at EW(Ly$\alpha$) $<10$\,\AA. Using the C\,\textsc{iii}] line peak as the systemic redshift, we calculate $\Delta v_{\rm{Ly}\alpha}$ and recover the previously reported trend of decreasing $\Delta v_{\rm{Ly}\alpha}$ with increasing EW(Ly$\alpha$). Interestingly, we find an anti-correlation between $\Delta v_{\rm{Ly}\alpha}$ and EW(C\,\textsc{iii}]), which also displays a dependence on the UV absolute magnitude. We derive a multi-variate fit to obtain $\Delta v_{\rm{Ly}\alpha}$ using both EW(C\,\textsc{iii}]) and $M_{\rm{UV}}$, finding that $\Delta v_{\rm{Ly}\alpha}$ is more strongly dependent on EW(C\,\textsc{iii}]), with a weaker but non-negligible dependence on $M_{\rm{UV}}$. We find that for a fixed EW(C\,\textsc{iii}]), UV-bright Ly$\alpha$ emitting galaxies show smaller values of $\Delta v_{\rm{Ly}\alpha}$,which suggests that such galaxies may be undergoing more bursty star-formation compared to the UV-fainter ones, akin to a population of extremely UV-bright galaxies identified at $z>10$.
\end{abstract}


\section{Introduction}

A largely uncharted period in the history of the Universe is the Epoch of Reionization (EoR), which began with the emergence of the first stars, black holes and galaxies after the Big Bang. It is widely believed that ultraviolet radiation from early sources governed the reionization of the intergalactic medium (IGM), which ended at a redshift $ z\sim 5.5-6$, \citep{fan06, bec13, mcg15, sal16, bos22}.

The duration of the EoR can be constrained by, amongst other probes, measures of the scattering and polarisation of the cosmic microwave background (CMB) \citep{hai99, zah12, benn13, pla16, cho20}, the opacity of hydrogen absorption in the spectra of distant quasars \citep{erb10, sta14, rig15, wei17, gar19, bos22}, and the visibility or otherwise of Lyman-$\alpha$ (\lya) emission in star-forming galaxies (SFGs) since this is resonantly scattered by neutral gas \citep{pen11, pri12, Jones23, Trapp23}. Constraining the sources responsible for reionization requires knowing their abundances and ionising capabilities.

With \emph{JWST}, at least constraining the average production rate of ionising photons in galaxies at $z>6$ is now within reach \citep{cam23, Tang23, katz23, curti23, Saxena23b}. However, as the neutral fraction of the IGM increases, it becomes impractical to directly observe the escaping ionising photons from galaxies beyond $z\simeq4$ \citep[e.g.][]{ino14}. Direct measures of the leakage of Lyman continuum (LyC) photons is only possible in lower redshift analogues, where such observations can be linked to other observables \citep[e.g.][]{fle19, izo21, sax22, flu22, pah23}. A further tool in understanding the physics of how LyC photons may escape is via high-resolution cosmological hydrodynamical simulations \citep[e.g.][]{maj22, cho23}.

\lya\ emission has historically been used to explore how ionising radiation can escape, since both \lya\ and LyC photons rely on relatively dust-free environments to escape \citep{ver17, izo21, naidu22, maj22, hay23}. The velocity offset of \lya\ with respect to the systemic velocity determined from other nebular lines in particular can trace the geometry of the distribution of neutral gas in an H\,\textsc{ii} region, with Ly$\alpha$ peaks close to the systemic velocity probing relatively low column densities that facilitate both the escape of \lya\ as well as LyC photons from galaxies \citep{dij06, dij14, ver15, ver17}.

Fundamentally, however, the increasing neutrality of the IGM renders \lya\ emission an unreliable tool in the EoR \citep[e.g.][]{sant04, dij06, hay10, beh13, ino14} and we must rely on other emission lines to infer the leakage of LyC photons. Rest-frame ultraviolet spectroscopy has revealed strong emission lines of the semi-forbidden \ciii\ $\lambda\lambda1907,1909$ doublet (referred to as \ciii\ hereafter) over $z\sim 2-7$ \citep{sha03, erb10, sta14, sta15} and various authors have discussed whether studies of \ciii\ emission may be a valuable probe as a substitute for \lya\ \citep{sta14, sta15, sta16, din17}. These observations however pre-date \emph{JWST}. With \emph{JWST} results delving deeper into the reionization era with spectroscopy, we are already seeing how \emph{JWST} is transforming identification of these emission lines \citep[e.g][]{Bunker23, Tang23, sax23}

Several studies have previously explored correlations between the equivalent widths (EW) of \lya\ and \ciii\ \citep{sha03, erb10, sta14, rig15, sta15, din17, sch18, hut19, lef19, Marchi19, Cull20, rav20, lle21} in low redshift SFGs where \lya\ is un-attenuated by the neutral IGM. \cite{Marchi19} found a strong correlation between the two lines, but noted that their sample may be biased as the sample was selected to have both \lya\ and \ciii\ emission, questioning whether this relation could change if the entire population of galaxies with \ciii\ emission were considered. The basic idea behind any possible correlation is that the conditions which support the efficient production and escape of \lya\ radiation (i.e. low dust content, low metallicity, partial coverage of neutral hydrogen) may be similar to those required for the production of collisionally-excited emission lines such as \ciii. However, \cite{rig15} pointed out that the correlation is driven primarily by the strongest emitters of \lya\ and \ciii\ (EW(\lya)\,$\sim 50$\AA  \:\& EW(\ciii)\,$\sim5$\AA\ ), whereas for weaker emitters there is no convincing correlation. Moreover, at low metallicities ($Z=0.1 - 0.2\,Z_\odot$), the correlation may weaken further \citep{nak17}. 

Regardless, any correlation between \lya\ and \ciii\ may offer the prospect of using \ciii\ emission from galaxies at $z>6$, now feasible with \emph{JWST}, to constrain the \lya\ properties and therefore the nature of the ionising sources and their environment. Using spectroscopic data from the VANDELS survey for SFGs in the redshift range $z\sim3-4$ where both \lya\ and \ciii\ are visible, we examine in further detail the utility of \ciii\ as a proxy for \lya.

The work explored in this paper complements earlier studies by \cite{Marchi19} and \cite{lle21} that have used VANDELS data to investigate these emission lines. \cite{lle21} explored the global properties of \ciii\ derived from stacked spectra within the VANDELS dataset. The unique aspect of this work is the emphasis on a larger sample of individual emitters compared to \cite{lle21} and, crucially, the offset between \ciii\ and \lya\ and its connection with Lyman continuum leakage. \cite{Marchi19} studied \ciii\ and \lya\ including offsets; however, a key difference in this work is that we do not pre-select for \lya.

The layout of this paper is as follows: In Section \ref{sec:data} we describe the VANDELS data used in this work and the various line measurement methods employed. We present the main results based on this spectroscopic data in Section \ref{sec:results}. Finally, we summarise the findings of this paper in Section \ref{sec:summary} and place the results into context with those in the literature and discuss the \ciii\ and \lya\ relationship along with the implications of the \lya\ offsets in the reionization era. 

Throughout the paper, we assume the following cosmology: $\Omega_m$ = 0.3, $\Omega_\Lambda$ = 0.7, H$_0 = 70$\,km\,s$^{-1}$\,Mpc$^{-1}$. We use the notation of a positive EW to imply emission, and all logarithms are in base 10 unless otherwise specified.

\section{Data and measurements}
\label{sec:data}

\begin{figure}
    \centering
    \includegraphics[width=\linewidth]{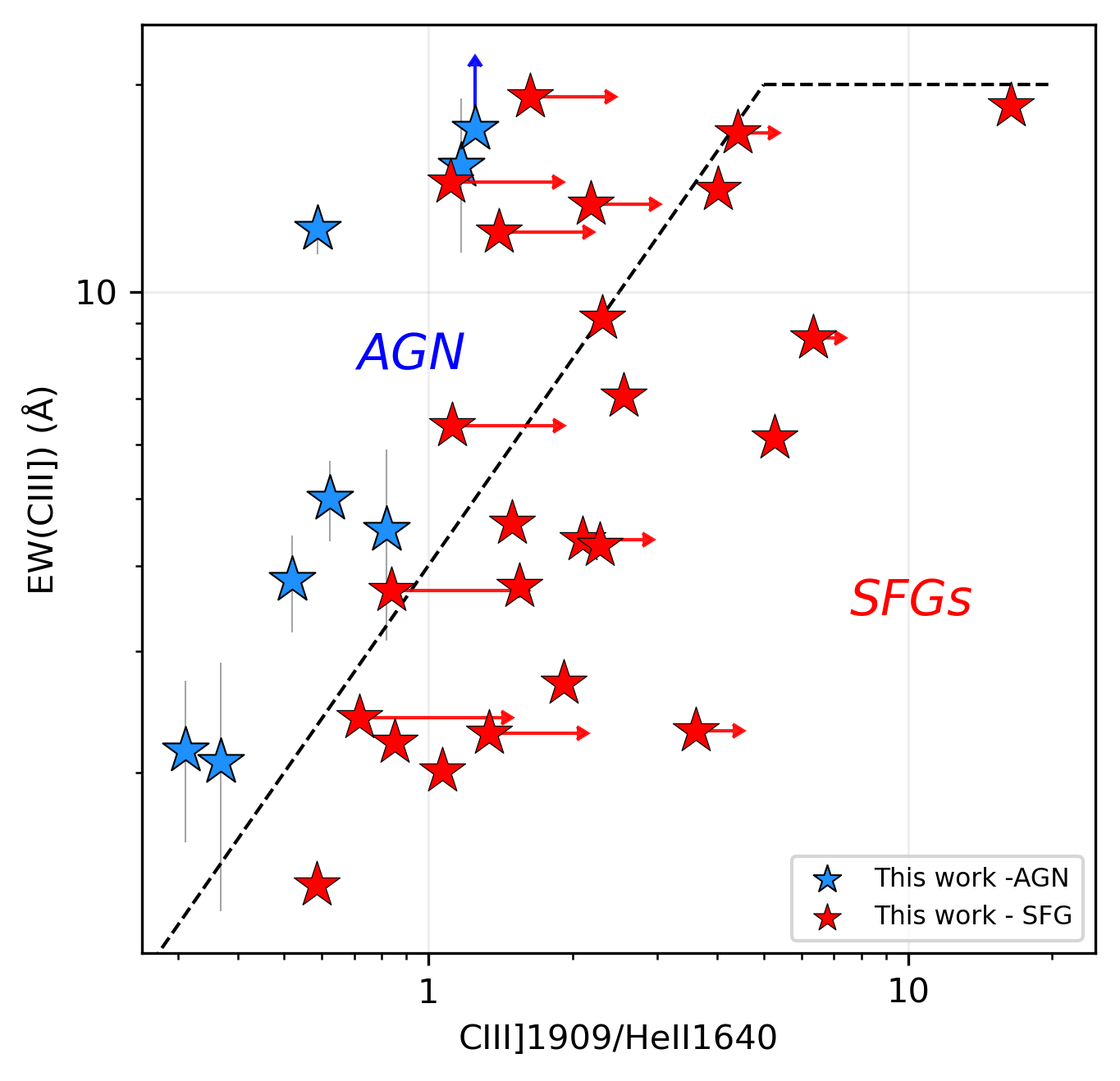}
    \caption{ \ciii/\heii\ flux ratio vs EW(\ciii) for CL3 \ciii\ and \lya\ emitting sources in our sample, with the dashed line representing the boundary between SFGs and AGN according to the photoionization models of \citet{nak17}. Red stars indicate those classified as SFG's according to the boundaries. The red stars with arrows are \ciii\ and \lya\ emitters with no \heii\ detection in their emission spectra and illustrate upper limits for those sources, which are consistent with photoionization via star formation. Blue stars are objects with \heii\ detections likely photoionized by AGN. Using this diagnostic (along with the CDFS deep X-ray catalogue, \citealt{Luo17}), we identify 9 objects from our sample of 139 CL3 \ciii\ and \lya\ emitters likely dominated by AGN, and remove them from our sample going forward.}
    \label{fig:nakajima-agn}
\end{figure}

\subsection{VANDELS survey}
The spectra used in this study all come from the VANDELS survey, which is a VIMOS survey of the CANDELS fields using the VIMOS spectrograph on ESO's Very Large Telescope (VLT) \citep{pen18, mcl18}. The survey covers a total area of $~0.2$ deg$^{2}$ centred on the CANDELS UKIDSS Ultra Deep Survey (UDS) and Chandra Deep Field South (CFDS) fields. It secured high signal-to-noise ratio (SNR) spectra at optical wavelengths $(0.48 < \lambda < 1.0 \mu$m) for $~2100$ galaxies in the redshift interval $1.0 \le z \le 7.0$ \citep{gar21}. Around 85\% of the galaxies targeted by VANDELS are at $z \sim 3$. The unique aspect of the VANDELS survey is that each source has a significant exposure time, ranging from 20hrs to a maximum of 80hrs \citep{pen18}. Full details and description of the survey and target selection can be found in \cite{mcl18}. 

\subsection{ \ciii\ and \lya\ emitter selection and line measurement}
\label{sec:measurements}
In this work, we focus on spectroscopically confirmed star-forming galaxies from VANDELS data release 4 (DR4) across both UDS and CDFS fields \citep{gar21} in the redshift range of 3 $\le z \le 4$, only selecting those galaxies that have a redshift probability of being 95\% correct (a redshift reliability flag of 3 or 4, see \citealt{pen18} for more details). This redshift range ensures that both the \ciii\ and \lya\ emission lines lie within the spectral range. These initial selection criteria resulted in a sample of 773 objects. 

To identify \ciii\ and \lya\ emission from this parent sample, both the 1D and 2D spectra were visually inspected using the \textsc{pandora} software \footnote{\url{https://www.iasf-milano.inaf.it/software/}}. Based on the robustness of the \ciii\ emission from visual inspection, the sample was split into three different groups based on the `confidence level' (CL hereafter). CL 3 was assigned to those sources with clear \ciii\ emission in both their 1D and 2D spectra, unaffected by sky features or residual noise. Spectra that showed clear \ciii\ emission in either 1D or 2D spectra (but not both) were assigned a CL of 2, whereas possible line detections likely contaminated by residual noise/sky features were assigned a CL of 1. Out of an initial parent sample of 773 galaxies, \ciii\ emission (CL 1-3) was identified in 280 galaxies, with 139 galaxies assigned the highest confidence level of CL3. 

As the main aim of this paper is to study the dependence of \ciii\ emission on \lya, we then proceeded to identify \lya\ emission from all those galaxies that had possible \ciii\ emission. We did not introduce a confidence level criterion for \lya\ emission, as the reliability and significance of any \lya\ emission was determined using emission line fitting (as discussed later). Across our parent sample, 62 galaxies were found to exhibit both \lya\ and \ciii\ (CL3) via visual inspection.

We then measured the strength of the emission lines in these 62 objects. The first step involved determining a `systemic' redshift using the \ciii\ line, for which we used a single Gaussian fit and the systemic redshift was determined from the peak of this Gaussian \citep[see][for example]{sax20}. Using the systemic redshift, each spectrum was de-redshifted and a single Gaussian was fitted to \lya, \ciii\ and all other rest-UV lines (such as \heii\ and C\,\textsc{iv}) visible in the spectra, with the local continuum value for each line estimated by polynomial fitting to nearby regions free from other emission or absorption features. EWs and integrated fluxes were determined from these Gaussian fits. If the S/N ratio of the continuum was below $2\sigma$, the $2\sigma$ noise level was used to determine a lower limit for the line EW.

Thanks to accurate systemic redshifts measured from the \ciii\ line, we could also accurately measure the \lya\ velocity offset from the systemic velocity. By comparing the centroid of the observed \lya\ emission with that expected for a rest-frame wavelength of 1215.67\AA\ using the galaxy systemic redshift, the velocity offset of the \lya\ peak was determined for all galaxies in our sample. The emission line fluxes, widths, EWs and the velocity offset of \lya\ compared to systemic redshift are listed in Table \ref{table:all data}.

\subsection{Identification of possible AGN}
Prior to further analysis, we considered it important to identify possible AGN in our sample to remove any biases. As \ciii\ has a high ionisation potential ($24.4$\,eV), it is possible that some of the stronger \ciii\ emitters are AGN. To explore this, we searched for another high ionisation line, \heii\ $\lambda 1640$ ($54.4$\,eV), which is also commonly seen in AGN. 

We then attempted to identify possible AGN using rest-UV line ratio diagnostics. Specifically, we used EW(\ciii) versus \ciii\,$\lambda 1909$/\heii\,$\lambda1640$ proposed by \cite{nak17}, which is shown in Figure \ref{fig:nakajima-agn}. This resulted in the identification of 9 possible AGN out of the 62 CL3 \ciii\ and \lya\ emitters, which were removed from our sample. Further, cross-matching with the deep X-ray deep catalogue that exists in CDFS \citep{Luo17} revealed one more source identified to be an AGN, which was also removed from our sample. We note that using the \ciii\ and \heii\ based AGN selection, we have reported 7 new possible AGN from VANDELS, which were not identified in previous studies \citep[e.g][]{sax20}. 

For those galaxies with no clear \heii\ detection, we estimated an upper flux limit by integrating the 1 $\sigma$ continuum-subtracted flux at the expected position (red stars with arrows in Figure \ref{fig:nakajima-agn}). Since these values were consistent for star-forming galaxies, they were retained in the sample. 

\subsection{Final sample}
\begin{figure}
   \centering
   \includegraphics[width=\linewidth]{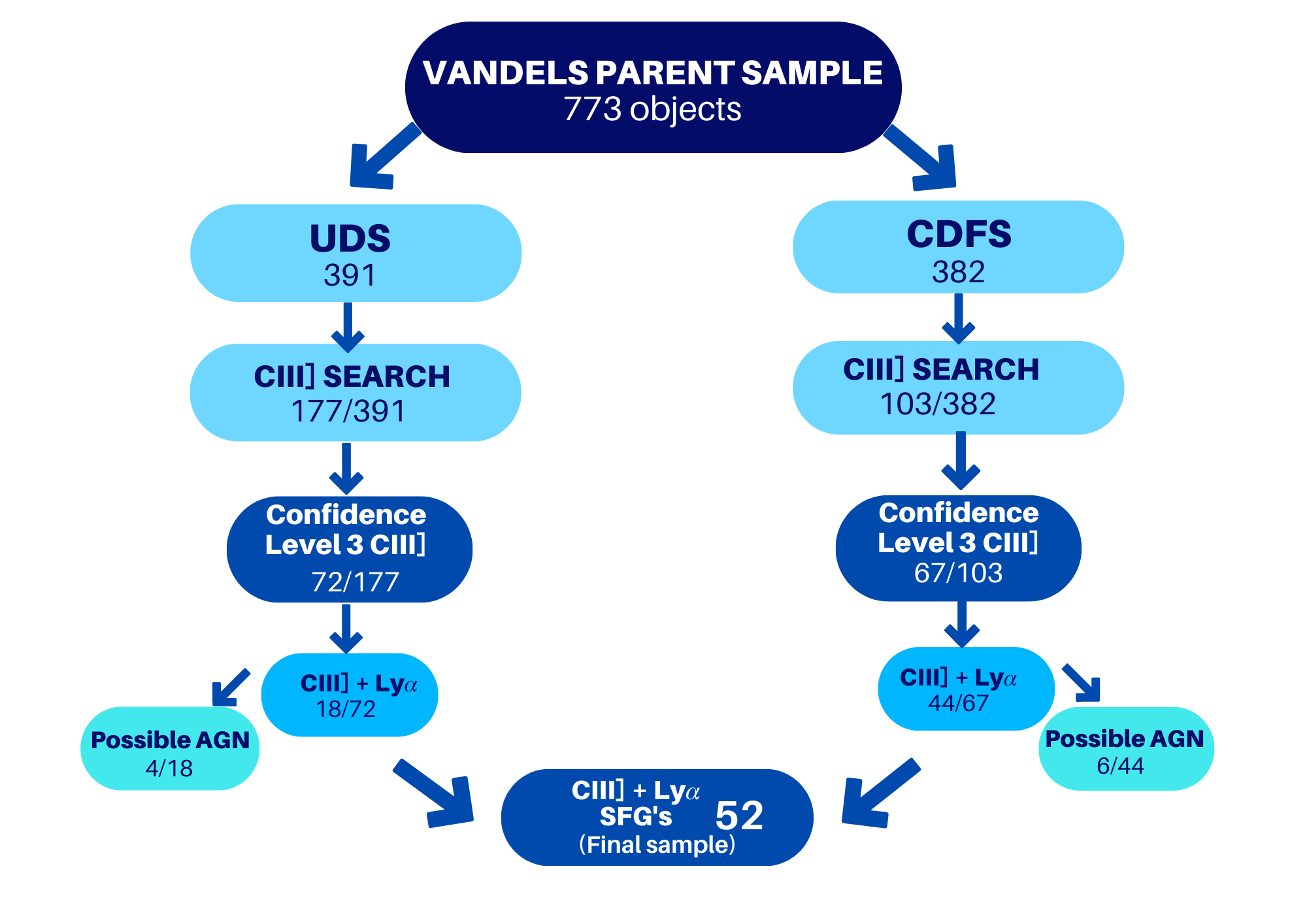}
    \caption{Flowchart showing the various samples and the number of objects retained after each stage of inspection/selection. Our final sample comprises 52 SFGs that are CL3 \ciii\ emitters with \lya\ emission and unlikely to host an AGN.}
    \label{fig:sample flowchart}
\end{figure}

The final sample of \ciii\ and \lya\ emitting SFGs in the redshift range $z\approx3-4$ comprises of 52 galaxies. Figure \ref{fig:sample flowchart} provides a breakdown of the overall sample and the various selection steps that were implemented to select the sources.

\begin{figure*}
   \centering
   \includegraphics[width=\linewidth]{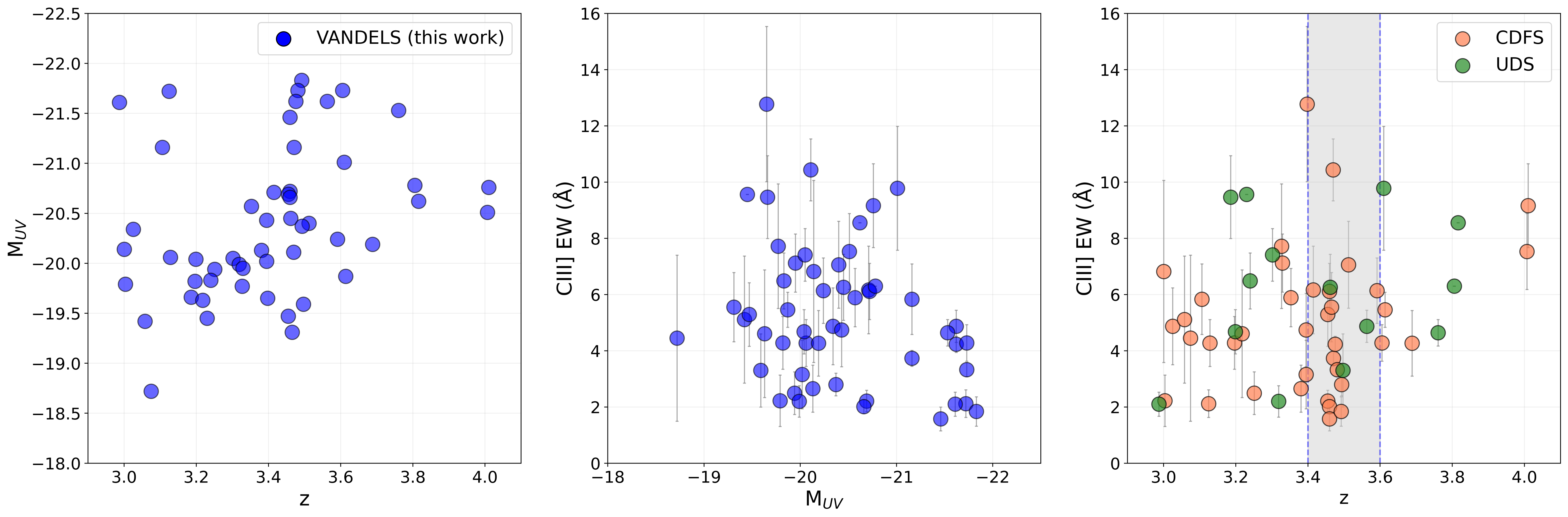}
   \caption{ This figure outlines the parameter space which our VANDELS sample occupies. \textit{(left):} Absolute UV magnitudes as a function of redshift for our final sample, demonstrating a wide range of $M_{\rm{UV}}$ covered by our sample as well as the incompleteness of fainter $M_{\rm{UV}}$ at high redshift. \textit{(middle):} Rest-frame EW(\ciii) as a function of $M_{\rm{UV}}$, showing a mild decrease in the EW(\ciii) at brighter $M_{\rm{UV}}$. \textit{(right):} The distribution of EW(\ciii) with redshift for sources in the UDS (green) and CDFS (orange) field. The shaded region highlights the known overdensity in the CDFS field at $z\sim3.5$, where we do find a slight excess of \ciii\ emitters.}
    \label{fig:3panelparam}
\end{figure*}

In Figure \ref{fig:3panelparam} we show the parameter space occupied by our \ciii\ emitter sample. In the left panel we show the distribution of $M_{\rm{UV}}$ and redshift for our final sample. As is evident from the Figure, our sample spans a relatively large range of UV magnitudes ($-22$ to $-18.5$), and the flux-limited nature of the survey naturally results in an incompleteness of fainter UV magnitudes at high redshifts. In the middle panel we show EW(\ciii) as a function of $M_{\rm{UV}}$, which shows a slight decrease in EW(\ciii) at brighter $M_{\rm{UV}}$. Finally in the right panel we show the EW(\ciii) as a function of redshift, which also highlights the known overdensity in CDFS at $z\sim3.5$ \citep{Skelton14, Straatman16, Forrest17, Luo17, Guaita20}. Apart from two objects with relatively high EW(\ciii) at $z\sim3.5$, we do not find any strong indication that there is an overall increase in the EW(\ciii) in the overdense region compared to the rest of the sample. There is also no appreciable clear trend between EW(\ciii) and redshift, although we acknowledge the possibility that \ciii\ detections may be influenced by sample incompleteness.

\section{Results}
\label{sec:results}

\subsection{Strength of \ciii\ emission}
Across our sample of 52 \ciii\ and \lya\ emitting galaxies, we measure an average EW(\ciii) of 5.0\,\AA. In Figure \ref{fig:3panelparam} we have highlighted the possible UV magnitude incompleteness resulting at higher redshifts because of the flux-limited nature of the sample. To address this, we divided our total sample of 52 galaxies into two sets based on $M_{\rm{UV}}$, with the `Bright' subsample with $M_{\rm{UV}}$ $<-20.4$ containing 24 galaxies and the `Faint' subsample with $M_{\rm{UV}}$ $\geq -20.4$ containing 28 galaxies.

The galaxies in the Bright subsample have EW(\ciii) in the range $1.6-9.8$\,\AA\ and an average EW(\ciii) of $5$\,\AA. Galaxies in the Faint subsample contains 28 galaxies showing a much larger range of EW(\ciii) in the range $2.2-12.8$\,\AA, with an average value of $5.6$\,\AA, which is marginally larger than that of the Bright subsample. These values along with those from other \ciii-selected samples in the literature are given in Table \ref{table: lit}. 

In Table \ref{table: lit}, we also compare our sample with similar samples targeting \ciii\ emitters available in the literature \citep{sha03, sta14, lef19, Cull20, lle21}. Our VANDELS sample represents a valuable increase in the total number of galaxies with both \ciii\ and \lya\ EWs over the full redshift range ($z\sim 0-10.6$) (as quoted in \cite{rav20}) from 59 to 111. Specifically, within the redshift range $z\sim3-4$, the sample increases from 38 to 80 galaxies with EW(\ciii)’s ranging from $1.6-12.8$\AA. Within our own sample we find that the UV-faint galaxies have a slightly higher average compared to the UV-bright galaxies.
\begin{table}\centering
  \centering
  \caption{Average EW(\ciii) and the range for \ciii\ emitters that also show \lya\ emission in this work and from other studies in the literature at redshifts $z \sim 2-5$ (\textit{Dashes represent no data publicly available)}.}
 
  \begin{tabular}{l*{3}{ccc r}}
  \toprule
   Data & EW(\ciii) [\AA] & Range [\AA] & N\\
    \midrule
    VANDELS (\textit{this work})  & 5.3      &  1.6 - 12.8     &  52 \\
    \midrule
    BRIGHT (\textit{this work})  & 5.0      &  1.6 - 9.8     &  24 \\
    FAINT (\textit{this work})  & 5.6      &  2.2 - 12.8     &  28 \\
    \midrule
    \cite{nak17}    & 2.0     &           -   &   -          \\ 
    \cite{sta14}    & 8.8   &  1.8 - 13.5  &    11        \\ 
    \cite{sha03}   & 3.6     &    1.9 - 5.7   &   2.0   \\
    \cite{rig15}   & 1.6     &     0.1 - 4.0  &      20        \\
    \cite{lef19} (\textit{stacked})    & 2.0 & -   &    -   \\ 
    \cite{lef19} (\textit{stacked})     & 2.2 & - &    -  \\ 
    \cite{lle21} (\textit{stacked})     & 3.9 & - &   - \\ 
    
  \bottomrule
  \end{tabular}
  \label{table: lit}
\end{table}

We note that although the Bright subsample is more complete across higher redshifts, the Faint subsample may be more representative of typical star-forming galaxies at $z>6$ that are now being routinely discovered by various \emph{JWST} surveys \citep{tacc22, End23, cam23, curti23, sax23}. Therefore, going forward, we colour code the VANDELS sample using their UV absolute magnitude when studying their \lya\ and \ciii\ properties where appropriate.

\subsection{Relationship between C\,\textsc{iii}] and \lya\ line strengths}
In this section we explore the relationships between \ciii\ and \lya\ emission in the VANDELS data from $3<z<4$ and compare our results to earlier studies at similar redshifts in the literature. We also discuss the dependence of \lya\ velocity offsets. 

\begin{figure}
   \centering
   \includegraphics[width=\linewidth]{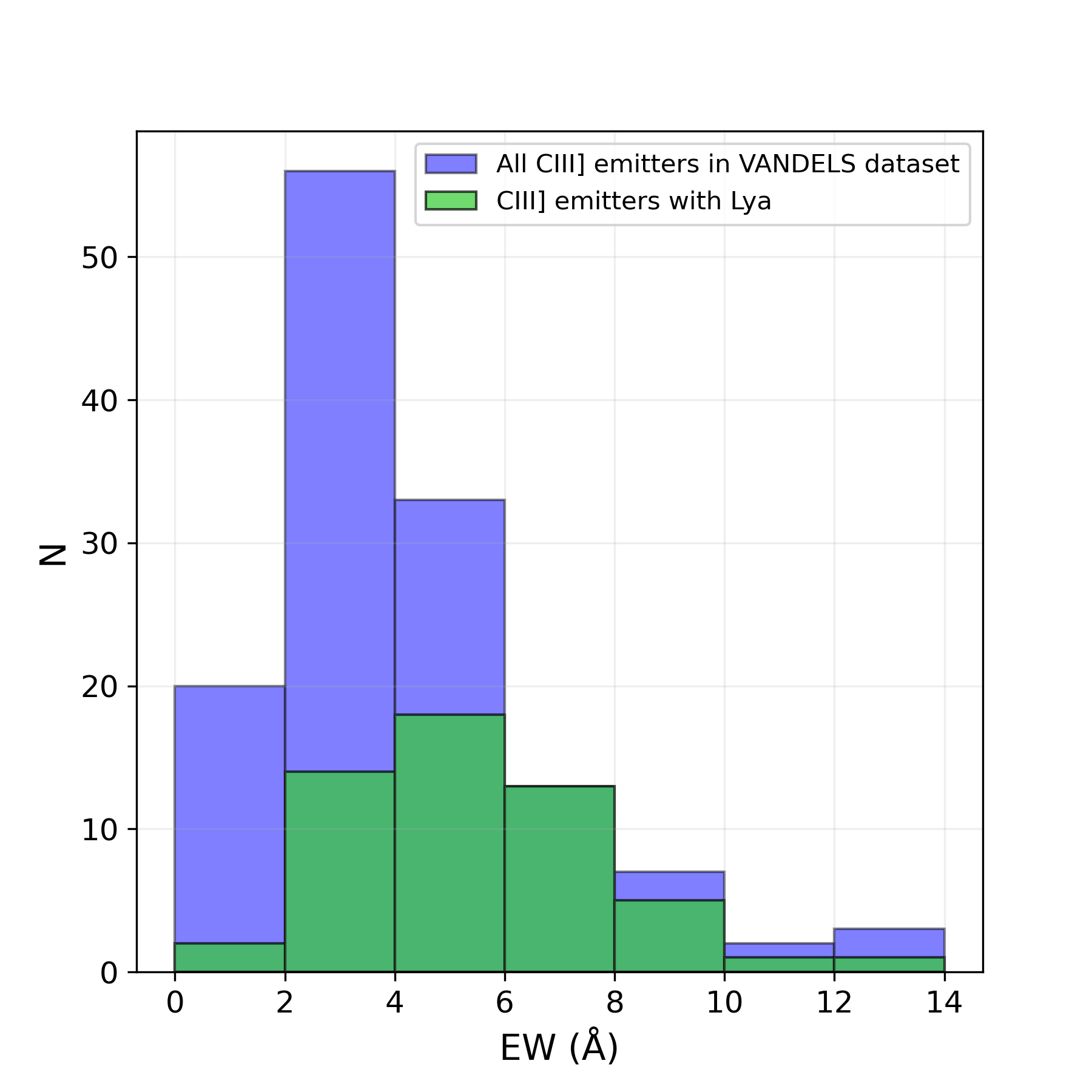}
   \caption{Histogram showing the distribution of EW(\ciii) for CL3 \ciii-emitters in our analysis. The blue histogram illustrates the full CL3 sample of \ciii, while the green illustrates the subsample of galaxies that have both \ciii\ and \lya\ emission. We find that \ciii\ emitters that also have \lya\ emission tend to have higher EW(\ciii) when compared to \ciii\ emitters that do not have strong \lya\ emission.}
    \label{fig:hist}
\end{figure}

Comparing with our original sample of only \ciii-selected galaxies from VANDELS, we find that \ciii\ emitters that show \lya\ emission have higher \ciii\ EWs overall compared to only \ciii\ emitters. This can be seen in Figure \ref{fig:hist} where we compare the histogram of the full CL3 \ciii\ emitters in the full VANDELS dataset (blue) and \ciii\ emitters with detectable \lya\ (green). The median \ciii\ EW of galaxies with both \ciii\ and \lya\ is $\sim 5$\,\AA ($\pm2.5$), compared to a median \ciii\ EW of only 2.9\,\AA\ ($\pm 4)$ for those across the full CL.3 sample \citep[see also][]{Cull20}. These \lya\ emitters having systematically higher EW(\ciii) may be explained by the presence of harder ionising radiation fields in such galaxies, tracing low metallicities and young stellar ages, as was also discussed by \citet{Cull20}. 

To explore any dependence of \ciii\ and \lya\ strengths on the absolute UV magnitudes of our final sample, we also investigate the line flux ratios between these two lines in both the Bright and Faint subsamples that we had created above. This reveals an average \lya/\ciii\ flux ratio of $8.42$ for Bright ($M_{\rm{UV}}$ < $-20.4$) galaxies, and a ratio of $5.99$ for Faint ($M_{\rm{UV}}$ > $-20.4$) galaxies in our final sample. This implies that, while possessing similar \lya\ strengths, UV-faint galaxies exhibit a more pronounced \ciii\ line intensity, once again potentially tracing the presence of younger stars and perhaps more `bursty' star-formation histories.

\begin{figure} %
    \centering
    {{\includegraphics[width=\linewidth]{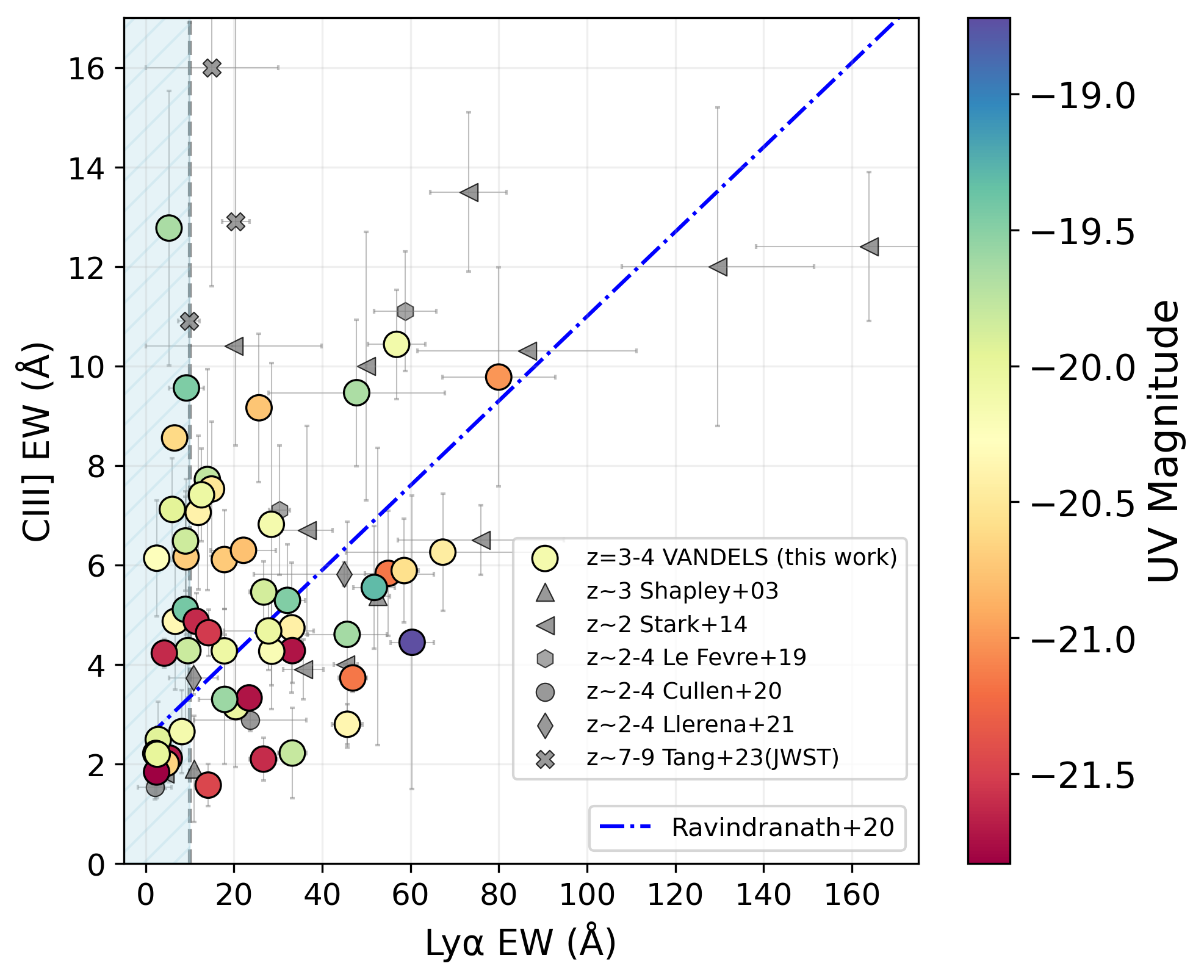} }}%
    \caption{Rest-frame EWs of \ciii\ and \lya\ in our VANDELS sample, colour coded by absolute UV magnitude, compared with samples from the literature spanning a redshift range $z \simeq$ 2-4 \protect\citep{sha03, sta14, lef19, Cull20, lle21}. We also include in the figure $z>7$ measurements from \textit{JWST}/NIRSpec from \protect\cite{Tang23} ($z \simeq$ 7-9). The blue dashed line shown represents the line of best fit outlined in \protect\cite{rav20} while the vertical grey line and shaded region presents a cut off at EW(\lya) $\le 10 \AA$ where our sample has a number of strong \ciii\ emitters with weak \lya\ emission.}
        \label{fig:ciii ew vs Lya ew}%
\end{figure}

To further explore correlations between \lya\ and \ciii\ strengths in our sample, in Figure \ref{fig:ciii ew vs Lya ew} we show the distribution of EW(\ciii) and EW(\lya), colour-coded by $M_{\rm{UV}}$. For demonstration purposes, we show the line of best fit as obtained by \cite{rav20} for a combined sample from the literature of \lya-selected star-forming galaxies at $z \simeq 0-7$ (depicted as the blue dashed line). From our sample alone, we do not find a clear correlation between EW(\ciii) and EW(\lya), which may be expected given how our sample has been selected. Compared to other literature samples, we have selected our galaxies based on \ciii\ detections alone. This means that there may be a number of \lya-emitting galaxies with no \ciii\ detection, which will be missed by our selection, rendering our sample relatively incomplete for any statistical analysis.

Interestingly however, at EW(\lya) $ <10$\,\AA, we observe a large scatter in EW(\ciii) shown via the grey dashed line and shaded region. The EW(\ciii) of galaxies in this regime ranges from $\sim1.6-13.0$\,\AA, which represents the widest range of EW(\ciii) for any given EW(\lya) bin. Large EW(\ciii) values often indicate conditions with high ionisation parameters and low metallicities, whereas \lya\ emission is affected by line-of-sight attenuation from regions of high neutral gas or dust. These effects may explain the low observed EW(\lya). Another possible explanation is offered by the presence of AGN powering the \ciii\ emission in this regime, which could perhaps have been missed by the UV line ratio diagnostics that we employed earlier.

We note that UV luminous galaxies in our sample (red points in Figure \ref{fig:ciii ew vs Lya ew}) occupy the lower part of the distribution, with low EW(\ciii) as well as low EW(\lya). These luminous galaxies are perhaps tracing more evolved, less star-forming systems in our sample, which would explain the weak EWs.

We further show \ciii\ detections reported from galaxies at $z>7$ from \cite{Tang23} using \textit{JWST}/NIRSpec spectroscopy. Interestingly, the galaxies from \citet{Tang23} also show very high EW(\ciii) values with low EW(\lya). At these redshifts, the low EW(\lya) can be explained by near complete attenuation of \lya\ emission by the intervening neutral IGM. This highlights that at the highest redshifts, we would expect to see strong \ciii\ even in the absence of strong \lya, as is the case for a handful of our VANDELS galaxies at intermediate redshifts.

\subsection{\lya\ velocity offset as a function of \lya\ equivalent width}
\begin{figure} %
    \centering
    \includegraphics[width=\linewidth]{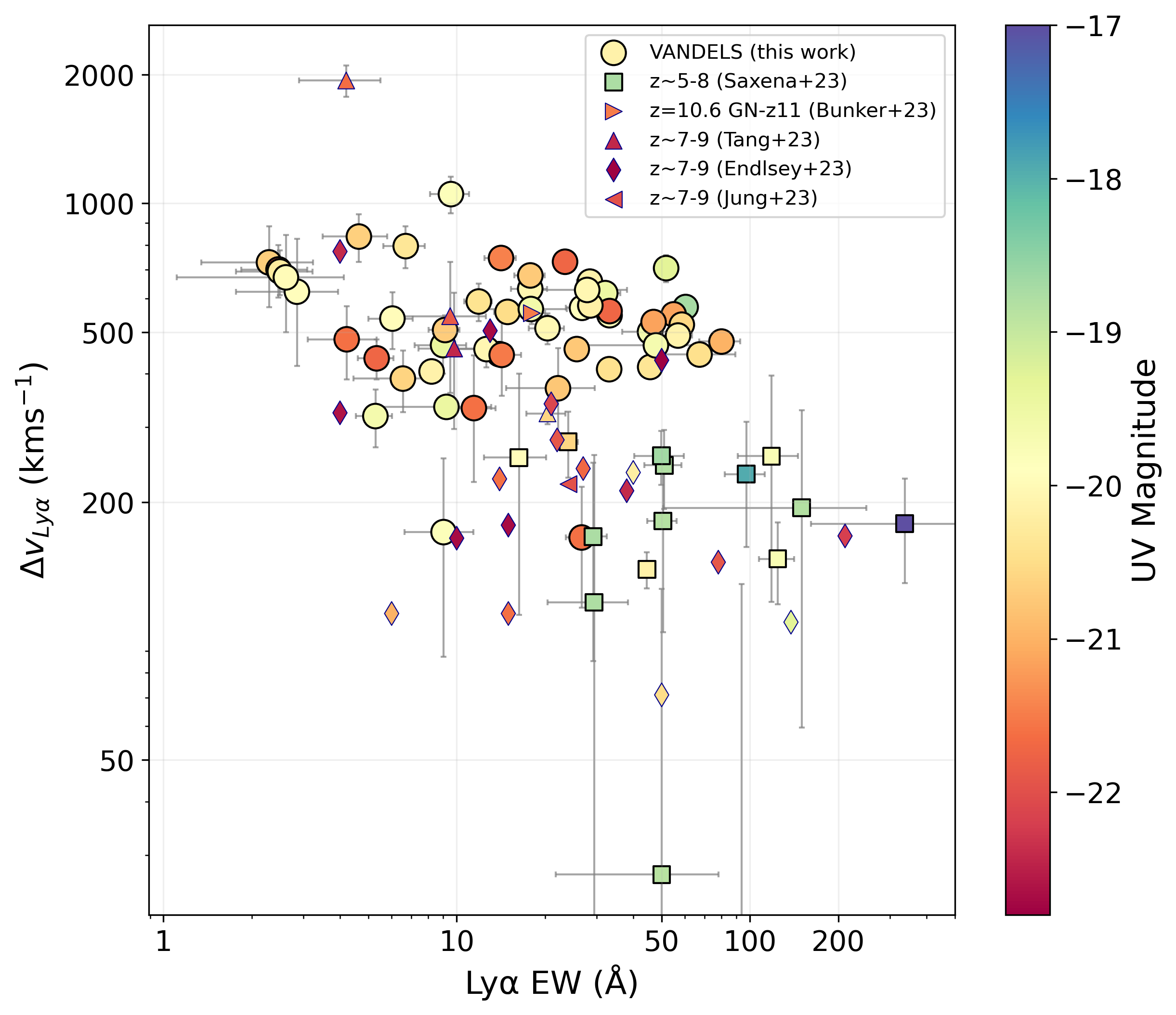}%
    \caption{\protect\lya\ EW vs \lya\ velocity offset for the VANDELS \ciii\ data in the context of \emph{JWST} literature \citep{Bunker23, Tang23, Saxena23b, End23, Jung23}, colour coded using $M_{\rm{UV}}$. Overall, we find a declining \lya\ velocity offset from systemic redshift with increasing \lya\ EWs, consistent with earlier work \citep[e.g.][]{erb10, Nakajima2018Ellis, Prieto23, Saxena23b}. We further note that UV-bright galaxies often tend to have weaker \lya\ EWs and higher velocity offsets.}
    
    \label{fig:lya vel off}%
\end{figure}

In this section we briefly explore the \lya\ properties of our galaxies. Across our sample, we measure \lya\ velocity offsets ranging from $166-1051$ km\,s$^{-1}$, with an average value of $\Delta v_{\mathrm{Ly}\alpha}$ $\simeq 533$\,km\,s$^{-1}$.  In Figure \ref{fig:lya vel off} we show the velocity offset from systematic redshift as a function of EW(\lya), colour-coded by $M_{\rm{UV}}$, finding qualitatively an anti-correlation between the \lya\ velocity offset and strength. This anti-correlation has also been previously reported in the literature for star-forming galaxies across redshifts \citep[e.g.][]{erb10, Nakajima2018Ellis, Tang23, Prieto23, roy23, Saxena23b}.

We also compare our \lya\ measurements with those available for a handful of reionization era \lya-emitting galaxies at $z>6$ that likely reside in ionized bubbles. Shown in the figure are \lya\ velocity offsets measured from LAEs as compiled by \citet{End23} using ground-based observations with systemic redshifts measured from other rest-UV lines or far-infrared lines from ALMA. Additionally, we show measurements from LAEs at $z>6$ from the latest \emph{JWST} surveys, which include the faint LAEs presented in \citet{Saxena23b} using deep spectroscopy from the JADES survey, and relatively brighter LAEs selected from CEERS \citep{Jung23, Tang23}. We additionally show the measurement from GN-z11 at $z=10.6$ from \citet{Bunker23}.

Overall, the $z>6$ LAEs agree with the general trend observed between EW(\lya) and $\Delta v_{\mathrm{Ly}\alpha}$ in our sample, and the scatter we find in our sample is highly consistent with other high redshift LAE observations \citep[e.g.][]{erb10, sta14, Willott_2015, Nakajima2018Ellis, cho20, Tang23, Prieto23}. This demonstrates that the $z>6$ LAEs must likely reside in patches of relatively ionized IGM, where the \lya\ attenuation and profile is largely being determined by scattering through the ISM. 

As has been noted by other studies, the anti-correlation between \lya\ velocity offset and EW is likely driven by the density and geometry of neutral gas in the systems, where a higher density leads to increased absorption/scattering of \lya\ photons along a line of sight, leading to a reduction in the observed EW and an increase in the velocity offset from systemic, which can be explained using the `expanding shell model' described in \cite{ver06, Dij16}. This explanation is supported by \cite{Marchi19}, who used a similar VANDELS dataset and found a correlation between the \lya\ velocity offset and the shift in the interstellar absorption lines, where galaxies with high ISM outflow velocities also showed small \lya\ velocity offsets. 

\cite{Prieto23} identified a notably reduced mean value of $\Delta v_{\mathrm{Ly}\alpha}$ at 205\kms\ for some of the faintest galaxies (avg. $M_{\rm{UV}}$ $\sim -17.8$) within a similar redshift range ($3<z<5$). This observation is expected since their sample primarily encompasses galaxies with limited diversity in terms of their stellar masses and consequently lower neutral gas densities, where a higher $\Delta v_{\mathrm{Ly}\alpha}$ may be less frequently observed.

Assuming that the observed \lya\ properties in our sample of galaxies are regulated by attenuation/scattering by neutral gas in the ISM at redshifts where the neutral IGM is not expected to attenuate \lya\ significantly, we can now attempt to understand the physical conditions that regulate both the strength and velocity profiles of \lya\ emission as well as the strength of \ciii\ emission. This can then be used to understand how \ciii\ emission may potentially be used as a diagnostic for \lya\ properties in reionization era galaxies that do not reside in ionized bubbles, and therefore do not show \lya\ emission.

\subsection{\ciii\ as a tracer of \lya\ velocity offset}
Having confirmed that the \lya\ properties are indeed sensitive to the neutral gas density that controls the escape of \lya\ photons along a line of sight, we now compare the velocity offset of \lya\ with the strength of \ciii\ emission to investigate whether the observed EW(\ciii) can be used as a diagnostic of neutral gas content, and consequently \lya\ and LyC escape from the ISM of high redshift galaxies. 

\begin{figure*} %
    \centering
    {{\includegraphics[width=14cm]{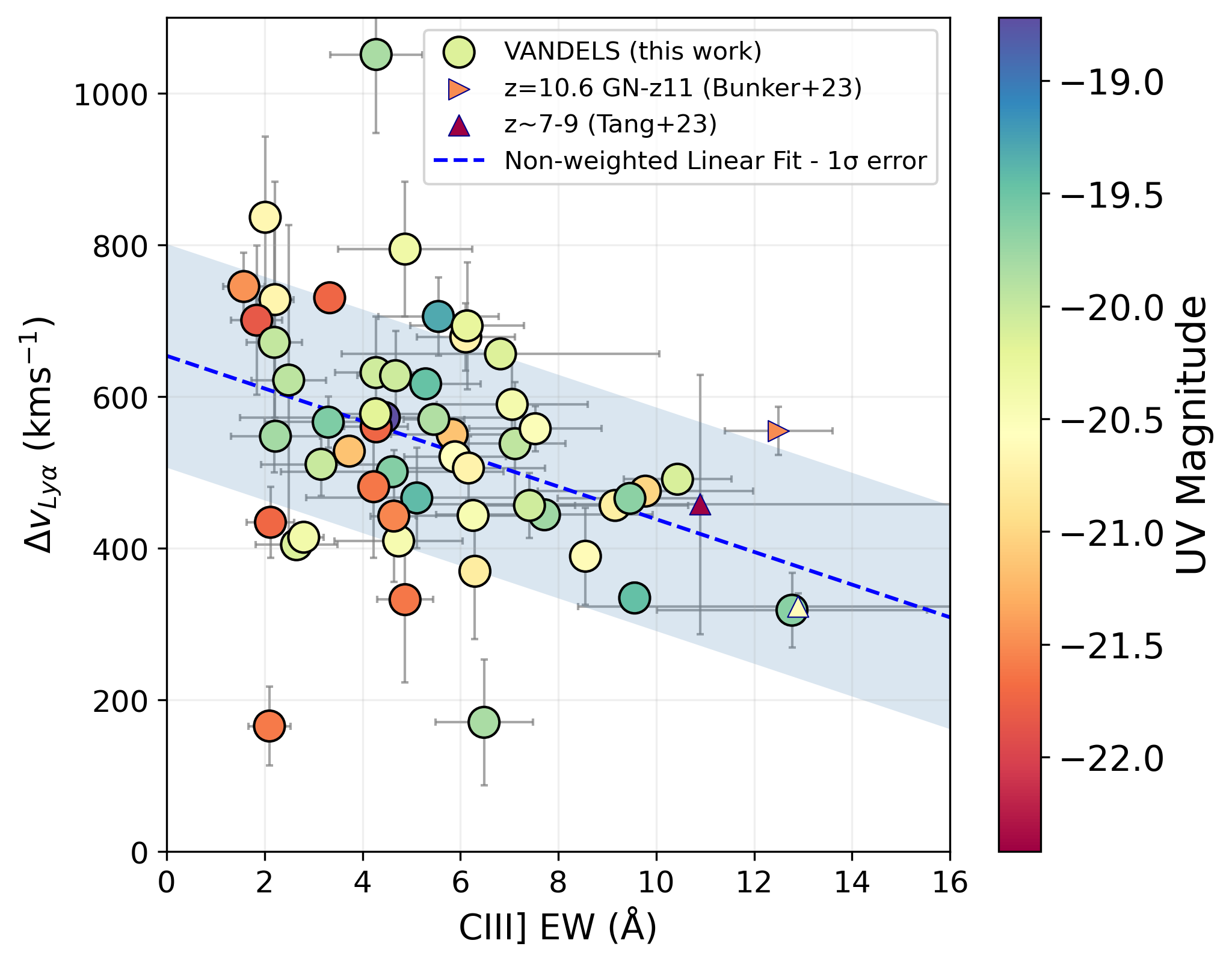} }}%
    \caption{\ciii\ EW vs \lya\ velocity offset for the VANDELS \ciii\ data, colour coded using $M_{\rm{UV}}$. The blue dashed line represents a non-weighted linear regression fit, and the shaded area represent the $1\sigma$ error for the VANDELS \ciii\ data. The bright UV galaxies have a larger scatter due to the increased gas compared to the faint UV galaxies.}
    \label{fig:ciii vel off}%
\end{figure*}

In Figure \ref{fig:ciii vel off} we show the distribution of \lya\ velocity offset as a function of EW(\ciii), once again colour-coded by $M_{\rm{UV}}$ for galaxies in our sample. Qualitatively speaking, there appears to be a tentative anti-correlation between the observed \lya\ offset and EW(\ciii), with UV-faint galaxies showing less scatter and UV-bright galaxies showing a larger scatter. Therefore, any possible correlation between \lya\ velocity offset and EW(\ciii) must take into account the UV absolute magnitudes which seem to affect the relation.

In light of this, we derive a correlation between \ciii\ and $\Delta v_{\mathrm{Ly}\alpha}$, taking into account $M_{\rm{UV}}$. To do this, we employ a multi-variate fit to quantify $\Delta v_{\mathrm{Ly}\alpha}$ as a function of both EW(\ciii) and $M_{\rm{UV}}$, standardising the independent variables ($x_i$), which takes the form:
\begin{equation}
    \frac{\Delta v_{\rm{Ly}\alpha} - \bar{{\Delta v}}_{\rm{Ly}\alpha}}{\sigma_{\Delta v}} = \sum_{i=1}^{2} A_i \frac{x_i - \bar{x_i}}{\sigma_i}
    \label{eq:ciii_vel_off}
\end{equation}
where $A_i$ are the coefficients for EW(\ciii) and $M_{\rm{UV}}$, $\bar{x_i}$ and $\sigma_i$ are the mean and standard deviation, respectively, of the independent variables. Performing a linear regression on this equation gives us the coefficients -0.388 for EW(\ciii) and 0.16 for $M_{\rm{UV}}$. The values of the coefficients as well as the mean and standard deviations of the independent variables are given in Table \ref{tab:coefficients}.
\begin{table}
    \centering
    \caption{Best-fitting coefficients to predict \lya\ velocity offset from the systemic redshift using EW(\ciii) and $M_{\rm{UV}}$.}
    \begin{tabular}{l c c c}
    \toprule
    Variable & $A_i$ & $\bar{x_i}$ & $\sigma_i$ \\
    \midrule
    ${\Delta v_{\rm{Ly}\alpha}}$  & & $539.5$ & $158.6$ \\
    \midrule
    EW(\ciii) & $-0.388$ & $5.3$ & $ 2.5$ \\
    $M_{\rm{UV}}$ & $+0.16$ & $-20.4 $ & $0.7 $ \\
    \bottomrule
    \end{tabular}
    \label{tab:coefficients}
\end{table}

The dashed blue line in Figure \ref{fig:ciii vel off} shows the relationship that we have derived between \lya\ velocity offset and EW(\ciii) only, which shows a more promising anti-correlation between the two quantities with a Spearman's rank correlation coefficient = -0.36 and p-value = 0.0008. The shaded region represents the $1\sigma$ uncertainty on this relation. The influence of $M_{\rm{UV}}$ on this relationship is evident in the Figure, with a notable split observed between the majority of UV luminous points falling below the trend line.

Interestingly, the $z>7$ \lya\ and \ciii\ emitting galaxies from \citet{Tang23} appear to line up well within the $1\sigma$ uncertainty on the relation that we have derived, as shown in Figure \ref{fig:ciii vel off}, suggesting the presence of large ionized bubbles around them. GN-z11 on the other hand seems to lie just outside the $1\sigma$ band. This may be consistent with the very low \lya\ escape fraction reported in \citet{Bunker23}, which in combination with a relatively large velocity offset of $\sim500$\,\kms\ suggests that GN-z11 is likely not situated within an ionized bubble \citep[see also][]{hay23}, and the emergent \lya\ has been heavily attenuated by the neutral IGM surrounding it. Based on the relations that we have derived, the \lya\ velocity offset emerging from the ISM will be predicted to be much lower (up to 300\,\kms\ lower) based on our multivariate regression.

We find that for a fixed $M_{\rm{UV}}$, an increase in EW(\ciii) results in a decreasing $\Delta v_{Ly\alpha}$. However, interestingly, at a fixed EW(\ciii) emission, UV-bright galaxies tend to exhibit a lower \lya\ velocity offset compared to UV-faint galaxies. Notably, as mentioned previously, the UV-bright galaxies tend to lie below the best-fit line we show in Figure \ref{fig:ciii vel off}. This behaviour suggests that for the same EW(\ciii), UV-bright galaxies may be in a more `bursty' phase of star-formation, potentially facilitating a more efficient clearing of channels through which \lya\ can escape. It has indeed been shown that galaxies in the early Universe may be expected to follow a more bursty star-formation that boosts their observed UV magnitudes \citep[e.g.][]{End23, Simmonds24}. Stronger outflows driven by a stronger starburst aligns with expectations from \lya\ escape in an expanding shell model \citep[e.g.][]{ver06, Dij16}, where neutral gas is expelled, allowing the \lya\ emission to escape close its systemic velocity.

Overall, our results indicate that the \ciii\ emission is likely originating from an ionising radiation field likely driven by young stars. In such a scenario, the UV magnitude serves as a measure of the number of stars undergoing formation. To validate this observation comprehensively, additional UV-bright galaxies with strong \ciii\ and \lya\ signals are required. \cite{erb14} also noted velocity offset to be correlated with UV luminosity, however found that their faint LAE selected sub-sample of galaxies had smaller \lya\ velocity offsets and did not see the correlations present in their UV-brighter comparison sample possibly due to lack of dynamical range in mass and luminosity.

In summary, our findings indicate that for a fixed $M_{\rm{UV}}$, a higher EW(\ciii) is associated with a lower velocity offset. Conversely, for a fixed EW(\ciii), a brighter $M_{\rm{UV}}$ results in a lower velocity offset. We propose that this effect may be contributing to the observed patterns. The observed inverse correlation between EW(\ciii), \lya\ velocity offset and its dependence on galaxy UV magnitude, therefore, promotes the utility of \ciii\ as a proxy for \lya\ from galaxies in the epoch of reionization. With increased \ciii\ line detections across samples of $z>6$ galaxies with NIRSpec, our derived correlations may lead to an additional albeit indirect estimate of the leakage of \lya\, and from that potentially the LyC photons from star-forming galaxies in the early Universe.

\section{Discussion and Summary}
\label{sec:summary}
By employing alternative indicators in the absence of direct \lya\ observations, we can still make significant strides in studying the processes that regulate the escape of \lya\ as well as LyC photons, essential to understand the key drivers of reionization. While \lya\ remains a crucial component for understanding reionization, its unavailability limits our capacity to obtain velocity offsets, which are crucial in estimating parameters such as \lya/Lyman continuum escape fractions and bubble sizes \citep{Prieto23, Saxena23b, Jones23, witstok23}.

Latest results constraining the \lya\ fractions and EW distributions at $z>6$ from \emph{JWST} have found the \lya\ fractions to be low, which is expected given the increasing neutral fraction of the IGM \citep[e.g.][]{Jones23}. This means that at the highest redshifts, unless star-forming galaxies are surrounded by ionized regions, observing \lya\ emission will become increasingly challenging. Particularly beyond $z>9.5$ as the prominent O and H lines shift beyond the detectable range of the \emph{JWST} NIRSpec, \ciii\ may emerge as a next best feature to both confirm redshifts and deduce additional properties. Therefore, we propose a potential solution to use the relatively bright \ciii\ emission line in rest-UV, coupled with $M_{\rm{UV}}$, as a tracer for \lya\ velocity offsets from systemic through the equation provided in Eq. \ref{eq:ciii_vel_off}.

With an increase in the number of detections of rest-UV lines at $z>8$ \citep{Tang23, Bunker23, Hsiao23, Fujimoto23, Topping24}, including the detection of \ciii\ emission at $z=12.5$ reported by \citet{Deugenio23}, it is becoming essential to maximize the utility of these emission lines in inferring all possible properties of galaxies and their surroundings. Our proposal to employ \ciii\ as an indirect tracer of \lya\ velocity offsets (Figure \ref{fig:ciii vel off}) represents an approach to utilise insights from this strong rest-UV line to better understand the escape of \lya\ photons from the ISM of galaxies in the reionization era, possibly evaluating their contribution towards reionizing their local bubble.

In this study, we have used the rest-UV spectra for a sample of star-forming galaxies in the redshift range $\approx3-4$ from the VANDELS survey with confirmed detections of \lya\ and \ciii\ lines, to explore the utility of using the \ciii\,$\lambda1909$ line, often the second brightest emission line after \lya\ in the rest-UV, to infer the intrinsic \lya\ properties of galaxies in the reionization epoch, where \lya\ is significantly attenuated by the neutral IGM. Starting with the detection of \ciii\ emission from 391 objects in the UDS and CDFS fields from the VANDELS survey, we find that 36\% of objects show \ciii\ emission, with an average EW $\sim$3\AA\ across the total \ciii\ population. Using the \ciii\ line as a prerequisite, we have then searched for \lya\ emission in the sample, recovering a final sample of 52 star-forming galaxies with robust detections of both \lya\ and \ciii\ in their spectra.

We find that the equivalent widths of \lya\ and \ciii\ are not significantly correlated, finding an increased scatter in EW(\ciii) for galaxies with particularly low EW(\lya) $< 10$\AA, in agreement with previous studies. We show that galaxies with strong \lya\ emission have a higher EW(\ciii) on average ($\sim$5\AA) compared to those without \lya, highlighting that \lya\ emitting galaxies that also show \ciii\ emission are likely tracing galaxies undergoing rapid star-formation. In this work we report some of the highest EW(\ciii) values (ranging from $1.6-12.8$\AA) when compared with those published in the literature at comparable redshifts with a range of $z\sim2-4$.

Using the peak of the \ciii\ emission as an indicator of the systemic redshift of galaxies in our sample, we calculate \lya\ velocity offsets, $\Delta v_{Ly\alpha}$, in the range $166-1051$\,km\,s$^{-1}$, with an average of $533$\,km\,s$^{-1}$. We find a weak anti-correlation between EW(\lya) and $\Delta v_{\rm{Ly}\alpha}$, as has been previously reported in the literature, attributed to the role of scattering of \lya\ photons by the neutral gas in the ISM.

Interestingly, we report an anti-correlation between EW(\ciii) and $\Delta v_{\rm{Ly} \alpha}$, which appears to also depend on the absolute UV magnitude of the galaxy. This dependence of EW(\ciii) on both $\Delta v_{\rm{Ly}\alpha}$ and $M_{\rm{UV}}$ is captured using a multi-variate equation, as shown in Equation \ref{eq:ciii_vel_off} and Table \ref{tab:coefficients}, where we find that $\Delta v_{\rm{Ly}\alpha}$ anti-correlates more strongly with EW(\ciii) (coefficient of -0.388), with a comparatively weaker but non-negligible correlation with $M_{\rm{UV}}$ (coefficient of +0.16).

From our multivariate equation, we find that for a fixed $M_{\rm{UV}}$, an increase in EW(\ciii) leads to a decrease in $\Delta v_{\rm{Ly}\alpha}$. At a fixed EW(\ciii), on the other hand, UV-bright galaxies show lower \lya\ velocity offsets compared to UV-faint galaxies. In our sample, we find that UV-bright galaxies tend to lie below the best-fitting relationship between only EW(\ciii) and $\Delta v_{\rm{Ly}\alpha}$, suggesting that UV-bright galaxies with strong \lya\ and \ciii\ emission may be undergoing a more `bursty' phase of star-formation, aiding the efficient clearing of channels through which \lya\ can escape, consistent with the `expanding shell model'. Although this is promising within the context of a population of extremely UV-bright galaxies that are now being discovered $z\gtrsim10$, to validate this trend, additional spectroscopy for galaxies at $z<6$ with comparably bright absolute UV magnitudes is essential.

Thanks to \emph{JWST} spectroscopy, it is now possible to detect \ciii\ emission out to $z\gtrsim10$, and therefore the relationship between $\Delta v_{\rm{Ly}\alpha}$ on EW(\ciii) and M$_{\rm{UV}}$ that we have presented in this work may provide insights into the nature of the intrinsic \lya\ emission that may be escaping from the ISM of these galaxies, before it encounters a relatively neutral IGM and gets attenuated along the line of sight. Obtaining a better handle on the intrinsic production and escape of \lya\ photons from galaxies in the reionization era can provide unprecedented insights into the dominant contributors of ionizing photon towards cosmic reionization.

\section*{Acknowledgements}
We thank the referee for a very constructive report that undoubtedly improved the quality of this manuscript. This work was supported by UK Research and Innovation (UKRI), Science and Technology Facilities Council (STFC). We would like to thank K.Nakajima for supplying values for the ionization model boundaries. RSE acknowledges financial support from European Research Council Advanced Grant FP7/669253. AS acknowledges funding from the ``First- Galaxies'' Advanced Grant from the European Research Council (ERC) under the European Union’s Horizon 2020 research and innovation programme (Grant agreement No. 789056). This work was made possible through the use of software packages including \textsc{astropy} \citep{AstropyCollaboration}, \textsc{matplotlib} \citep{Matplotlib}, \textsc{mpdaf} \citep{mpdaf}, \textsc{pandas} \citep{reback2020pandas} and \textsc{topcat} \citep{topcat}.

\section*{Data Availability}
The data underlying this paper are publicly available from the VANDELS data release page \url{http://vandels.inaf.it/dr4.html} and from the European Southern Observatory (ESO) archive \url{http://archive.eso.org/cms.html}. The \textsc{python} code used to analyze the data will be made available upon request.


\bibliographystyle{mnras}
\bibliography{mnras_ciii.bib}


\appendix
\section{Data table}
\onecolumn
\begin{longtable}{l c | c c c | c c c c c}
\caption{Line measurements for confidence level 3 (CL3) \ciii\ and \lya\ emitters in the VANDELS sample.} \label{table:all data} \\
\toprule
Object ID & $z_\mathrm{spec}$ & \multicolumn{3}{c}{\ciii\ $\lambda1909$} & & \multicolumn{3}{c}{Ly$\alpha\,\lambda1216$} & \\
& & EW$_0$ & FWHM & Flux & EW$_0$ & FWHM & Flux & $\Delta v_{\mathrm{Ly}\alpha}$ & $M_{\rm{UV}}$ \\
& & (\AA) &  (km\,s$^{-1}$)   & $\times10^{-18}$  & (\AA)   & (km\,s$^{-1}$)   & $\times10^{-18}$ &  (km\,s$^{-1}$)  \\
      &   &  &  & (erg\,s$^{-1}$\,cm$^{-2}$)  &  &  & (erg\,s$^{-1}$\,cm$^{-2}$) &   \\
\midrule
\normalfont{CDFS}\\
\midrule
\endfirsthead
\multicolumn{10}{c}{{Table \thetable{} continued from previous page}} \\
\toprule
Object ID & $z_\mathrm{spec}$ & \multicolumn{3}{c}{\ciii\ $\lambda1909$} & & \multicolumn{3}{c}{Ly$\alpha\,\lambda1216$} & \\
& & EW$_0$ & FWHM & Flux & EW$_0$ & FWHM & Flux & $\Delta v_{\mathrm{Ly}\alpha}$ & $M_{\rm{UV}}$ \\
\midrule
\endhead
\bottomrule
\multicolumn{10}{r}{{Continued on next page}} \\
\endfoot
\bottomrule
\endlastfoot
\normalfont{CDFS226606} &	4.0102&	9.1 $\pm$ 1.5  &	678&	1.28 $\pm$ 0.07 &	25.7 $\pm$ 1.7 &	894&	5.69 $\pm$ 0.22 &	456$\pm$ 17 &	-20.76 \\
\normalfont{CDFS231741} &	4.0068&	7.5 $\pm$ 1.4  &	671&	0.71 $\pm$ 0.08 &	14.9 $\pm$ 1.5 &	1039&	1.98 $\pm$ 0.13 &	558$\pm$ 30 &	-20.51\\
\normalfont{CDFS226868} &	3.6884&	4.3 $\pm$ 1.2 &	  490&	0.36 $\pm$ 0.09 &	28.6  $\pm$ 1.2 &	913&	3.57 $\pm$ 0.05 &	577 $\pm$ 10 &	-20.19 \\
\normalfont{CDFS000303} &	3.6137&	5.5 $\pm$ 0.6 &	  583&	0.22 $\pm$ 0.02 &	26.8  $\pm$ 3.1 &	832&	1.78 $\pm$ 0.01 &	570 $\pm$ 7 &	-19.87 \\
\normalfont{CDFS017345} &	3.6052&	4.28 $\pm$ 0.6 &	863&	1.28 $\pm$ 0.2 &	33.4 $\pm$ 1.8 &	815&	22.1 $\pm$ 0.85 &	560$\pm$ 12 &	-21.73\\
\normalfont{CDFS233973} &	3.5913&	6.1 $\pm$ 1.2 &	  983&	0.89 $\pm$ 0.12 &	2.5  $\pm$ 0.7 &	484&	0.53 $\pm$ 0.15 &	693 $\pm$ 84 &	-20.24\\
\normalfont{CDFS015347} &	3.5124&	7.1 $\pm$ 1.5  &	927&	0.81 $\pm$ 0.1 &	11.9 $\pm$ 1.3 &	1257&	1.85 $\pm$ 0.18 &	590$\pm$ 59 &	-20.40 \\
\normalfont{CDFS020583} &	3.4932&	2.8 $\pm$ 0.4 &	  552&	0.30 $\pm$ 0.04 &	45.7  $\pm$ 3.5 &	787&	8.49 $\pm$ 0.16 &	414 $\pm$ 7 &	-20.37 \\
\normalfont{CDFS122764} &	3.492&	1.8 $\pm$ 0.5  &	625&	0.86 $\pm$ 0.05 &	2.5 $\pm$ 0.6 &	553&	1.87 $\pm$ 0.43 &	700$\pm$ 99 &	-21.83\\
\normalfont{CDFS128455} &	3.4813&	3.3 $\pm$ 0.3  &	734&	1.00 $\pm$ 0.06 &	23.4 $\pm$ 1.8 &	837&	17.9 $\pm$ 0.88 &	730$\pm$ 20 &	-21.73\\
\normalfont{CDFS004717} &	3.4756&	4.2 $\pm$ 0.3  &	737&	1.35 $\pm$ 0.07 &	4.2 $\pm$ 1.1 &	619&	2.21 $\pm$ 0.56 &	481$\pm$ 94 &	-21.62\\
\normalfont{CDFS140458} &	3.4708&	3.7 $\pm$ 0.3  &	659&	0.9 $\pm$ 0.03 &	46.9 $\pm$ 3.2 &	993&	16.2 $\pm$ 0.33 &	528$\pm$ 10 &	-21.16 \\
\normalfont{CDFS009692} &	3.4698&	10.4 $\pm$ 1.1 &	885&	0.85 $\pm$ 0.02 &	56.8  $\pm$ 6.6 &	890&	9.27 $\pm$ 0.19 &	491 $\pm$ 10 &	-20.11\\
\normalfont{CDFS012637} &	3.4655&	5.6 $\pm$ 1.2 &	  611&	0.22 $\pm$ 0.03 &	51.8  $\pm$ 4.7 &	1609&	3.10 $\pm$ 0.22 &	706 $\pm$ 52 &	-19.31 \\
\normalfont{CDFS105360} &	3.4598&	1.6 $\pm$ 0.4  &	416&	0.54 $\pm$ 0.14 &	14.2 $\pm$ 1.7 &	756&	8.79 $\pm$ 0.74 &	745$\pm$ 44 &	-21.46\\
\normalfont{CDFS006417} &	3.4597&	2.0 $\pm$ 0.2  &	706&0.33 $\pm$ 0.04 &	4.6 $\pm$ 1.2 &	667&	1.03 $\pm$ 0.25 &	837$\pm$ 106 &	-20.66\\
\normalfont{CDFS012448} &	3.4596&	6.1 $\pm$ 1.0  &	464&1.04 $\pm$ 0.09 &	17.8 $\pm$ 2.1 &	1293&	4.88 $\pm$ 0.30 &	679$\pm$ 44 &	-20.72 \\
\normalfont{CDFS103010} &	3.455&	2.2 $\pm$ 0.4  &	468&0.40 $\pm$ 0.05 &	2.3 $\pm$ 0.9 &	524&	0.49 $\pm$ 0.18 &	728$\pm$ 155 &	-20.69\\
\normalfont{CDFS208906} &	3.4548&	5.3 $\pm$ 1.1 &	   708&0.54 $\pm$ 0.07 &	32.1  $\pm$ 4.1 &	835&	5.68 $\pm$ 0.12 &	617 $\pm$ 7 &	-19.47 \\
\normalfont{CDFS246390} &	3.4149&	6.2 $\pm$ 1.6  &	596&0.97 $\pm$ 0.07 &	9.1 $\pm$ 1.1 &	884&	2.64 $\pm$ 0.24 &	506$\pm$ 42 &	-20.71\\
\normalfont{CDFS225147} &	3.3976&	12.8 $\pm$ 2.8 &	564&0.61 $\pm$ 0.05 &	5.3  $\pm$ 0.7 &	597&	0.58 $\pm$ 0.06 &	318 $\pm$ 49 &	-19.66 \\
\normalfont{CDFS212043} &	3.395&	4.7 $\pm$ 1.3  &	708&0.79 $\pm$ 0.11 &	33.1 $\pm$ 2.8 &	885&	9.06 $\pm$ 0.15 &	410$\pm$ 7 &	-20.43 \\
\normalfont{CDFS215423} &	3.395&	3.2 $\pm$ 1.2 &	   313&	0.51 $\pm$ 0.13 &	20.4  $\pm$ 2.8 &	1181&	3.63 $\pm$ 0.26 &	511 $\pm$ 42 &	-20.02\\
\normalfont{CDFS206968} &	3.3808&	2.7 $\pm$ 0.8 &	  369&	0.24 $\pm$ 0.07 &	8.2  $\pm$ 0.5 &	813&	1.45 $\pm$ 0.01 &	405 $\pm$ 15 &	-20.13 \\
\normalfont{CDFS024919} &	3.3528&	5.9 $\pm$ 1.0 &	  486&	0.85 $\pm$ 0.09 &	58.6 $\pm$ 3.3 &	857&	15.9 $\pm$ 0.22 &	520$\pm$ 5 &	-20.57\\
\normalfont{CDFS229681} &	3.3291&	7.1 $\pm$ 1.0 &	  596&	0.78 $\pm$ 0.04 &	6.0  $\pm$ 1.0 &	907&	1.15 $\pm$ 0.16 &	538 $\pm$ 81 &	-19.95\\
\normalfont{CDFS217955} &	3.327&	7.7 $\pm$ 2.2 &	   752&	0.72 $\pm$ 0.11 &	14.0  $\pm$ 1.3 &	908&	2.04 $\pm$ 0.07 &	444 $\pm$ 15 &	-19.77 \\
\normalfont{CDFS209992} &	3.251&	2.5 $\pm$ 0.8 &	   426&	0.27 $\pm$ 0.04 &	2.9  $\pm$ 1.1 &	612&	0.41 $\pm$ 0.14 &	622 $\pm$ 205 &	-19.94 \\
\normalfont{CDFS221849} &	3.2177&	4.6 $\pm$ 2.3 &	  630&	0.42 $\pm$ 0.03 &	45.6  $\pm$ 9.0 &	965&	5.69 $\pm$ 0.11 &	501 $\pm$ 10 &	-19.63\\
\normalfont{CDFS000598} &	3.1965&	4.3 $\pm$ 0.9 &	  576&	0.44 $\pm$ 0.04 &	9.6  $\pm$ 1.4 &	1002&	0.89 $\pm$ 0.10 &	1051 $\pm$ 104 &	-19.82 \\
\normalfont{CDFS007847} &	3.1284&	4.3 $\pm$ 0.8 &	  480&	0.58 $\pm$ 0.05 &	17.8  $\pm$ 2.5 &	1237&	2.72 $\pm$ 0.32 &	632 $\pm$ 74 &	-20.06 \\
\normalfont{CDFS113988} &	3.1246&	2.1 $\pm$ 0.5  &  628&	1.25 $\pm$ 0.28 &	5.3 $\pm$ 0.79 &	783&	6.12 $\pm$ 0.70 &	434$\pm$ 47 &	-21.72\\
\normalfont{CDFS023527} &	3.106&	5.8 $\pm$ 1.3  &682&	1.95 $\pm$ 0.08 &	54.9 $\pm$ 5.2 &	1124&	30.5 $\pm$ 1.23 &	550$\pm$ 20 &	-21.16\\
\normalfont{CDFS032490} &	3.0747&	4.4 $\pm$ 3.0 &	715&	0.14 $\pm$ 0.01 &	60.4  $\pm$ 4.9 &	947&	2.92 $\pm$ 0.09 &	573 $\pm$ 15 &	-18.72 \\
\normalfont{CDFS206035} &	3.0578&	5.1 $\pm$ 2.3 &	809&	0.66 $\pm$ 0.10 &	9.0  $\pm$ 1.8 &	664&	1.19 $\pm$ 0.22 &	466 $\pm$ 67 &	-19.42 \\
\normalfont{CDFS003496} &	3.0253&	4.9 $\pm$ 1.4 &	713&	0.90 $\pm$ 0.10 &	6.7  $\pm$ 1.1 &	842&	1.7 $\pm$ 0.26 &	795 $\pm$ 89 &	-20.34\\
\normalfont{CDFS015428} &	3.0037&	2.2 $\pm$ 0.9 &	878&	0.29 $\pm$ 0.03 &	33.2  $\pm$ 3.2 &	797&	5.24 $\pm$ 0.13 &	548 $\pm$ 10 &	-19.79\\
\normalfont{CDFS022563} &	3.0004&	6.8 $\pm$ 3.2 &	694&	0.85 $\pm$ 0.06 &	28.5  $\pm$ 2.7 &	903&	8.05 $\pm$ 0.27 &	656 $\pm$ 19 &	-20.14\\
    
    \midrule
    \normalfont{UDS}\\
    \midrule
\normalfont{UDS026657} \dag &	3.8161&	<8.5 &	980&	0.88 $\pm$ 0.09 &	6.6 $\pm$ 2.1 &	531&	1.51 $\pm$ 0.17 &	389$\pm$ 64 &	-20.62\\
\normalfont{UDS003783} \dag &	3.8056&	<6.3  &	836&	0.72 $\pm$ 0.15&	22.1 $\pm$ 7.4  &	768&	4.39 $\pm$ 0.43 &	369 $\pm$ 89 &	-20. 78\\
\normalfont{UDS295074} &	3.7606&	4.6 $\pm$ 0.5  &	1020&	1.02 $\pm$ 0.08 &	14.2  $\pm$ 2.3  &	735&	6.67 $\pm$ 0.66 &	443 $\pm$ 87 &	-21.53\\
\normalfont{UDS015507} &	3.61&	9.8 $\pm$ 2.2 &	1224&	1.65 $\pm$ 0.10 &	79.9 $\pm$ 12.8 &	841&	24.9 $\pm$ 0.50 &	475  $\pm$ 19 &	-21.01 \\
\normalfont{UDS151133} &	3.5628&	4.9 $\pm$ 0.6  &	908&	1.5 $\pm$ 0.13 &	11.5 $\pm$ 2.1  &	756&	7.50 $\pm$ 0.91 &	332 $\pm$ 109 &	-21.62 \\
\normalfont{UDS026015} &	3.497&	3.3 $\pm$ 1.3 &	371&	0.19 $\pm$ 0.06 &	17.9  $\pm$ 5.8 &	827&	1.47 $\pm$ 0.05 &	566 $\pm$ 33 &	-19.59 \\
\normalfont{UDS003724} &	3.4616&	6.3 $\pm$ 1.3    &	793&	0.74 $\pm$ 0.10  &	67.3 $\pm$  21.8  &	1242&	13.7 $\pm$ 0.17 &	443 $\pm$ 12 &	-20.45 \\
\normalfont{UDS376176} &	3.3188&	2.2 $\pm$  0.6 &	404&	0.25 $\pm$ 0.06 &	2.6 $\pm$  1.5  &	514&	0.36 $\pm$ 0.08  &	672 $\pm$ 172 &	-19.99 \\
\normalfont{UDS020394} &	3.3018&	7.4 $\pm$ 0.9 &	698&	0.67 $\pm$ 0.06&	12.6  $\pm$ 2.5 &	823&	2.19 $\pm$ 0.09 &	456 $\pm$ 42 &	-20.05\\
\normalfont{UDS000166} &	3.24&	6.5 $\pm$ 0.9  &	489&	0.51 $\pm$ 0.05&	9.0 $\pm$ 2.4 &	614&	0.67 $\pm$ 0.08 &	170 $\pm$ 83 &	-19.83 \\
\normalfont{UDS380039} \dag &	3.23&	<9.6   &	649&	0.78 $\pm$ 0.18 &	9.2 $\pm$ 3.9 &	744&	2.26 $\pm$ 0.06 &	334 $\pm$ 20 &	-19.45\\
\normalfont{UDS026615} &	3.199&	4.7 $\pm$ 0.8 &	739&	0.50 $\pm$ 0.07 &	27.8  $\pm$ 10.2 &	915&	6.0 $\pm$ 0.33 &	628 $\pm$ 58 &	-20.04\\
\normalfont{UDS019280} &	3.186&	9.5 $\pm$ 1.5 &	841&	1.03 $\pm$ 0.13 &	47.8  $\pm$ 19.9  &	945&	7.06 $\pm$ 0.10 &	466  $\pm$ 16 &	-19.66 \\
\normalfont{UDS197598} &	3.000&	2.1 $\pm$  0.4 &	827&	1.34 $\pm$ 0.24 &	26.7 $\pm$ 3.1  &	516&	40.7 $\pm$ 1.48 &	166 $\pm$ 52 &	-21.61\\

\midrule
\midrule
  \textit{Average ($1\sigma$)} & & 5.3 ($\pm2.5$) &  680 ($\pm187 $) & 0.76 ($\pm$ 0.39) & 17.8 ($\pm19.6$) &  836 ($\pm 226$) & 3.34 ($\pm 8.15$) & 533 ($\pm 158$) \\
  \midrule
  \midrule
  \normalfont{AGN}\\
  \midrule
  \normalfont{CDFS006905} * & 3.6885  & 6.5 $\pm$ 1.5 &  749  & 0.85 $\pm$ 0.16 &  16.7 $\pm$ 1.9  &  2206  & 2.91 $\pm$ 0.31 & 442 $\pm$ 12 \\
\normalfont{CDFS006327} * & 3.4937  & 4.5 $\pm$ 1.4 &  678  & 0.19 $\pm$ 0.02 &  26.2 $\pm$ 2.4  &  976  & 1.39 $\pm$ 0.39 & 606 $\pm$ 12\\
  \normalfont{CDFS025897} * & 3.4706  & 5.0 $\pm$ 0.7 &  929  & 0.84 $\pm$ 0.08 &  13.6 $\pm$ 1.5  &  732  & 3.80 $\pm$ 0.22 & 522 $\pm$ 12 \\
  \normalfont{CDFS217560} * & 3.4535  & 2.1 $\pm$ 0.8 &  534  & 0.19 $\pm$ 0.07 &  56.8 $\pm$ 6.9  &  827  & 8.52 $\pm$ 0.12 & 409 $\pm$ 12 \\
  \normalfont{CDFS247279} * & 3.2959  & 14.1 $\pm$ 3.6 & 650  & 1.1 $\pm$ 0.05 &  7.3 $\pm$ 1.6  &  1161  & 1.49 $\pm$ 0.31 & 916 $\pm$ 96 \\
  \normalfont{CDFS020988} * & 3.2638  & 3.8 $\pm$ 0.6 &  486  & 0.09 $\pm$ 0.01 &  51.7 $\pm$ 10.5  &  921  & 2.66 $\pm$ 0.05 & 498 $\pm$ 12 \\
\normalfont{UDS300247}  * & 3.794 &  15.3 $\pm$ 3.9 &  923  &  1.03 $\pm$ 0.14 & 22.2 $\pm$ 4.0 &  778  & 5.24 $\pm$ 0.19 & 452 $\pm$ 30 \\
 \normalfont{UDS006692}  * & 3.7663 &  2.1  $\pm$ 0.5 &   471 & 0.34 $\pm$ 0.08 & 7.3 $\pm$ 1.2  &  748  & 2.03 $\pm$ 0.28 & 455 $\pm$ 132 \\
  \normalfont{UDS382631} *\dag & 3.2176  & <17.2  &  1178  & 2.45 $\pm$ 0.17 &  60.1 $\pm$ 6.9  &  826  & 29.4 $\pm$ 0.40 & 453 $\pm$ 12 \\
  \normalfont{UDS145830} * & 3.2094 &  12.4  $\pm$ 1.0 &  1118  &  1.16 $\pm$ 0.06&  81.2  $\pm$ 13.3 &  983  & 7.51 $\pm$ 0.99 & 66 $\pm$ 15 \\

\bottomrule
\end{longtable}
\begin{tablenotes}
\item \textbf{Note:} '*' denotes an AGN candidate that was removed in the final sample and not included in the quoted average values. 
\item '$\dag$' denotes those objects with a lower limit for \ciii.
\end{tablenotes}

\bsp	
\label{lastpage}
\end{document}